\documentclass[10pt, conference]{IEEEtran}

\usepackage[square,numbers,sort&compress]{natbib}
\usepackage{flushend}
\usepackage{tabularx, color, colortbl}
\usepackage{graphicx, subfigure}
\usepackage{amsmath, amssymb}
\usepackage{algorithm}
\usepackage{algpseudocode}
\usepackage{listings}
\usepackage[noblocks]{authblk}
\usepackage{mathtools}
\usepackage{todonotes}
\usepackage{framed}
\usepackage{subfig}
\usepackage{soul}
\usepackage{verbatim}
\usepackage{url}
\usepackage{booktabs}

\newcommand*\circled[1]{\tikz[baseline=(char.base)]{
\node[shape=circle,fill,inner sep=1pt] (char) {\textcolor{white}{#1}};}}

\begin{document}
\title{AdapTBF: Decentralized Bandwidth Control via Adaptive Token Borrowing for HPC Storage}

\author{
    Md Hasanur Rashid\IEEEauthorrefmark{1}, 
    Dong Dai\IEEEauthorrefmark{1}
    \\
    \IEEEauthorrefmark{1}Department of Computer and Information Sciences, University of Delaware, Newark, US.
    \\
    mrashid@udel.edu,
    dai@udel.edu
}

\maketitle
\IEEEpubidadjcol

\begin{abstract}
Modern high-performance computing (HPC) applications run exclusively on computational resources but share global storage systems. This design can lead to issues when applications use a disproportional amount of storage resources compared to their allocated computing resources. An application running on a single compute node might consume excessive I/O bandwidth from a storage server, for example, by issuing numerous small, random writes. In doing so, it can hinder larger jobs that also write to the same storage server and are allocated many compute nodes, resulting in significant resource waste.

A straightforward solution to prevent such an issue is to limit each application’s I/O bandwidth on storage servers in proportion to its allocated computation resources. This approach has been implemented in existing parallel file systems using the Token Bucket Filter (TBF) mechanism. However, applying such methods in practice often results in lower overall I/O efficiency. HPC applications are known for generating short, bursty I/O requests. Therefore, strictly limiting I/O bandwidth proportionally either leads to wasted storage server bandwidth when applications are not performing I/O, or it prevents applications from temporarily utilizing higher bandwidth during bursty I/O phases.

We argue that the goal of I/O control should be to maximize both the I/O performance of each application and the overall efficiency of the storage server, while ensuring fairness among jobs, such as preventing smaller jobs from blocking large-scale ones. In this paper, we propose AdapTBF to achieve this objective. Building on the well-established TBF mechanism in modern parallel file systems (e.g., Lustre), AdapTBF introduces a decentralized bandwidth control approach using an adaptive borrowing and lending mechanism. We detail the algorithm, implement AdapTBF on the Lustre file system, and evaluate it using synthetic workloads modeled after real-world scenarios to demonstrate its effectiveness. Experimental results show that AdapTBF effectively manages I/O bandwidth for applications on storage servers while maintaining high overall storage utilization, even under extreme conditions.
\end{abstract}

\section{Introduction}
\label{sec:introduction}

Data-intensive scientific applications increasingly consider I/O performance a critical concern in high-performance computing (HPC) platforms~\cite{hey2009fourth}. Although new storage technologies (e.g.,
NVM SSD~\cite{welch2013optimizing,islam2023dgap}), complex storage tiers (e.g., burst buffer~\cite{liu2012role}), and various I/O optimization strategies~\cite{rashid2023iopathtune,egersdoerfer2024ion,dai16graphmeta} have steadily enhanced overall HPC I/O bandwidth, I/O contention and interference among concurrent applications in shared HPC storage still affect the performance of individual applications~\cite{lofstead2010managing,egersdoerfer2024understanding}. 
When the affected applications have already been allocated substantial computational resources, this interference can significantly reduce overall system utilization. 
In practice, it is not uncommon to see a small job congests the entire storage system by issuing excessive I/O requests (i.e., bandwidth hogging). To prevent such deliberate or unintended workloads from consuming a disproportionate share of I/O resources and hindering other applications, HPC storage systems must ensure that I/O bandwidth is allocated proportionally based on observed I/O demands and allocated computational resources.

Guarding I/O bandwidth for applications fundamentally involves monitoring and controlling their I/O activities. Extensive research has been conducted in this area~\cite{dai2014two,huang2003stonehenge,wang2012cake,thereska2011sierra,gulati2009parda,wu2007providing,li2016pslo,qian2017configurable}. Some of the earlier approaches employed fair sharing queuing techniques like SFQ~\cite{goyal1996start} and its extensions~\cite{jin2004interposed,wang2007proportional,xu2012vpfs} to proportionally allocate I/O bandwidth across different streams. However, in highly concurrent environments like distributed and parallel storage systems, SFQ can lead to unfair scheduling and prolonged wait times~\cite{jin2004interposed}.

Modern HPC parallel file systems instead typically utilize token bucket methods to manage the I/O bandwidth of various applications~\cite{thereska2013ioflow,stefanovici2016sroute,qian2017configurable}. For instance, Lustre employs the Token Bucket Filter (TBF) mechanism to regulate I/O bandwidth and prioritize access to the Lustre file system~\cite{braam2019lustre}. TBF operates by associating each client or user with a ``bucket” filled with tokens that represent the permission to execute I/O operations. Tokens are consumed as data is read or written, and the bucket is refilled at a predefined rate.

However, standard token bucket methods are not work-conserving, which can result in low storage server utilization during bursty I/O patterns, a typical behavior of many HPC applications~\cite{lofstead2011six,carns2011understanding}. 
Burst I/Os require the token bucket filter to dynamically adapt to fluctuations in I/O activity. 
When applications are not I/O-intensive, tokens should not be wasted, and storage servers should not remain idle. Conversely, when applications generate bursty I/O, more tokens should be allocated—if there is no significant competition for storage resources—allowing applications to utilize the available bandwidth proportionally to their needs.

\begin{figure*}[htpb]
\begin{center}
\includegraphics[width=\linewidth]{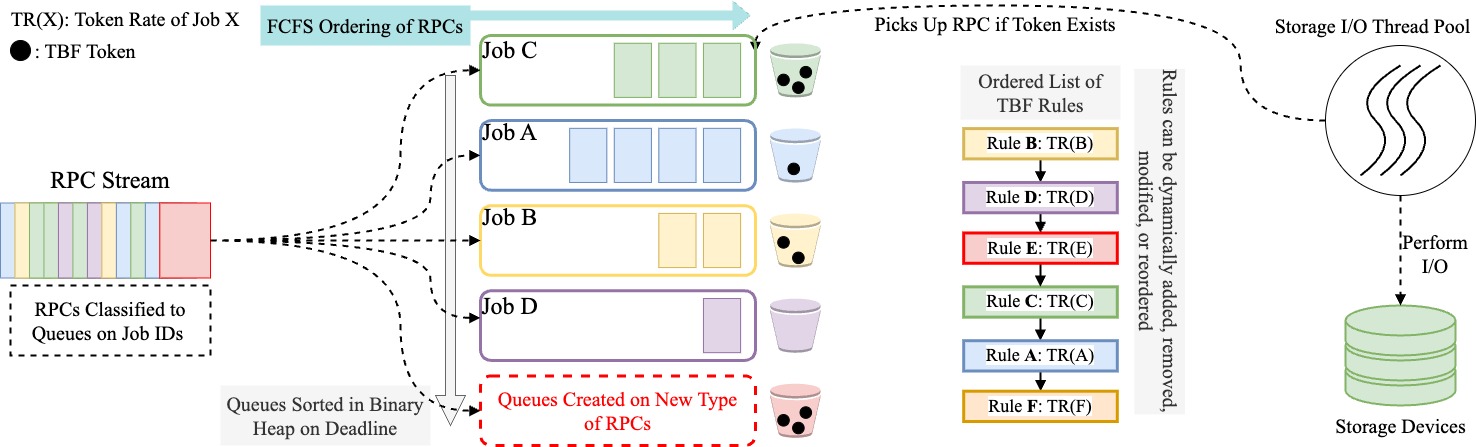}
    \caption{Lustre Token Bucket Filter (TBF) Mechanism.}
\label{fig:tbf}
\end{center}
\end{figure*}

We argue that the goal of guarding I/O bandwidth in HPC platforms is not solely about enforcing strict proportional bandwidth limits for each application. Equally important is maximizing the overall performance and utilization of the storage servers. For example, an application should be allowed to use all available I/O bandwidth when no other applications are actively performing I/O, even if this exceeds the proportion it is allocated. Similarly, if an application is unable to fully utilize its proportional bandwidth due to inefficient I/O patterns (e.g., small random I/Os), the remaining bandwidth should be made available to other applications adaptively to optimize the use of the storage system.

In this study, we introduce AdapTBF, a decentralized bandwidth control mechanism based on the Token Bucket Filter (TBF) framework, designed to manage and safeguard the I/O bandwidth of HPC applications on each storage server. The core objective of AdapTBF is to ensure that an application’s I/O bandwidth is proportional to its allocated computational resources (e.g., compute nodes) on each storage server, while simultaneously maintaining high I/O efficiency and maximizing the utilization of the storage server.

The key innovation of AdapTBF is its adaptive token-borrowing mechanism, which ensures per-application I/O bandwidth is managed in a work-conserving manner. We developed a prototype of AdapTBF based on Lustre file system, leveraging its token bucket filter mechanism. Extensive evaluations were conducted on CloudLab clusters~\cite{duplyakin2019design} using synthetic I/O workloads derived from real-world scenarios. The results show that AdapTBF effectively safeguards I/O bandwidth, ensuring appropriate bandwidth shares where feasible, while maintaining high utilization of the overall storage system. The main contributions of this research are two-fold:

\begin{itemize}
\item Propose a new adaptive token-borrowing mechanism to extend the existing token bucket filter framework, enabling local I/O bandwidth control in a work-conserving manner.
\item Implement the adaptive token-borrowing mechanism on the Lustre file system and evaluate its efficiency across a wide range of scenarios.
\end{itemize}

The rest of this paper is organized as follows: Section~\ref{sec:background}
presents the background and challenges for this research study. We introduce TBF mechanism in Lustre and its issues.
Section~\ref{sec:design} introduces the detailed design and implementation of
AdapTBF. Section~\ref{sec:evaluation} describes the evaluation and results
of this study. We conclude our study and discuss possible future
work in Section~\ref{sec:conclusion}.

\section{Background and Challenges}
\label{sec:background}

\begin{figure*}[htpb]
\begin{center}
\includegraphics[width=\linewidth]{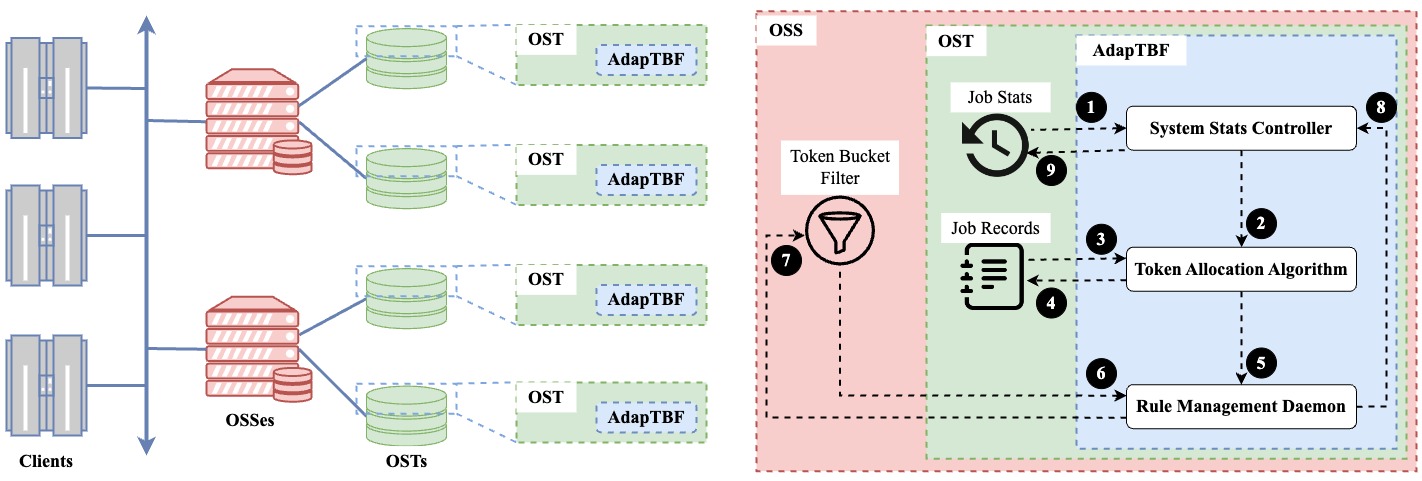}
    \caption{AdapTBF Framework Architecture. \textit{Left shows the overall architecture of HPC storage system. AdapTBF works on each object storage target (OST) to control I/O bandwidth in a decentralized way.}}
\label{fig:arch}
\end{center}
\end{figure*}

\subsection{TBF Mechanism in Lustre}

The Token Bucket Filter (TBF) policy~\cite{qian2017configurable} is a rate control mechanism used within the Lustre Network Request Scheduler (NRS) to enforce Quality of Service (QoS) by regulating the rate of Remote Procedure Calls (RPCs). In large-scale parallel file systems like Lustre, multiple clients and jobs compete for shared I/O resources, making it crucial to manage how RPC requests are handled to prevent resource starvation and maintain fairness. The TBF policy introduces a structured approach to enforce strict limit on RPC rates based on predefined rules, ensuring that each queue or job receives appropriate access to I/O resources.

In TBF, each queue represents a set of RPC requests grouped based on specific attributes, such as network ID (NID), job ID (JobID), and others. The queues are created according to predefined TBF rules when new RPC requests arrive. In Figure~\ref{fig:tbf}, we show the example of queues created based on JobID. Whenever a new RPC request arrives, the NRS TBF scheduler checks whether the RPC matches any existing rules. If matched, the RPC is placed in the corresponding queue for that rule. If no matching rule is found, the scheduler assigns the RPC to a fallback queue.

The TBF rules are maintained in an ordered list independently of the queues, allowing rules to be dynamically added, modified, reordered, or removed at runtime. Each queue linked to a rule is assigned a token bucket, which regulates the rate at which RPCs are processed. The bucket accumulates tokens at a rate specified by the rules. However, each bucket has a maximum capacity (e.g., 3 tokens by default), meaning it cannot hold more than this number of tokens at any time. If the bucket reaches its maximum capacity, any excess tokens are discarded to prevent a sudden burst of RPCs from overloading the system. For queues that do not have an associated rule, their RPCs are processed opportunistically whenever an idle I/O thread is available, and these queues are not subject to any token rate limits.

RPC requests within each queue are processed in a First-Come, First-Serve (FCFS) manner and will only be dequeued if enough tokens are available. The deadlines for each queue represent the time required to accumulate enough tokens to process an RPC. 
Queues are organized in a binary heap based on their deadlines, ensuring the system always prioritizes the queue with the nearest deadline. After processing, the system updates the queue’s token count and sets a new deadline according to the rate. This approach guarantees fairness across queues, as each queue is given the opportunity to process RPCs according to its rate.

\subsection{Challenges and Key Design of AdapTBF}

In practice, HPC applications can run on tens of thousands of compute nodes, issuing millions of I/O requests concurrently. Controlling the I/O bandwidth of these applications would require aggregating the I/O activities of all processes across multiple compute nodes before applying bandwidth control, which introduces significant scalability challenges. 

Instead of tracking and controlling I/Os on compute nodes, the approach can leverage storage servers, as prior work has shown~\cite{qian2017configurable,thereska2013ioflow}, to differentiate and monitor I/O activities from different applications. This is feasible as the number of storage servers, typically in the hundreds, is far smaller than the number of processes, which can reach the tens of thousands. However, even with hundreds of storage servers, globally monitoring I/O activities and making control decisions remains problematic due to the high concurrency and variability in I/O requests. This challenge motivates the design of a decentralized control mechanism, as described in Section~\ref{sec:design}.

The decentralized I/O control allows us to operate on the local conditions of each individual storage server. We argue this would work because if bandwidth sharing at the local storage level is both fair in allocation and efficient in utilization, the cumulative effect across all storage servers (which may contain multiple storage targets) will result in a globally fair allocation for active applications and ensure efficient resource utilization throughout the entire system.

Nevertheless, controlling application bandwidth on each storage target remains challenging due to the bursty nature of HPC I/O operations and the frequent changes in I/O access patterns during runtime. Additionally, the set of active applications on each storage server is highly dynamic, making it impractical to assume a static set of applications for bandwidth control. Manually allocating the correct portion of bandwidth to applications in a timely and appropriate manner becomes exceedingly difficult, if not impossible.

The goal of managing I/O bandwidth in HPC extends beyond simply enforcing a proportional share of resources among applications. Another critical objective is to maximize the utilization of the storage server itself. This work-conserving requirement ensures that if an application does not fully utilize its allocated bandwidth, the spare bandwidth should be made available to other applications immediately, rather than being reserved or left idle. To achieve this, I/O control must act swiftly to prevent throttling other applications. However, when spare bandwidth is allocated to other applications, it is equally important to ensure that the original application can reclaim its share if its I/O load increases, thereby maintaining long-term fairness.

\section{AdapTBF Design and Implementation}
\label{sec:design}

\subsection{Overall Design}
In this study, we propose AdapTBF, a framework to adaptively control I/O bandwidth on the shared storage servers utilizing the Lustre Token Bucket Filter (TBF) mechanism. The AdapTBF framework operates independently across all shared storage servers in a decentralized manner, relying solely on local metrics. It automatically determines and adjusts the token rate for jobs based on their allocated computing resources and actual I/O demands, aiming to maximize both job and storage efficiency together. AdapTBF leverages our new token lending and borrowing mechanism while ensuring fairness for jobs participating in token exchanges.

As illustrated in Figure~\ref{fig:arch}, the AdapTBF framework is implemented on top of the Lustre file system. The left side of the figure shows the typical layout of an HPC system using Lustre, where a large number of clients issue I/O requests to shared storage servers (OSS) over a shared network. Each shared storage server consists of one or more storage targets (OSTs). The AdapTBF framework operates independently for each Object Storage Target (OST), utilizing Lustre’s TBF mechanism located in the Object Storage Server (OSS) layer to adjust token rates.

AdapTBF operates as described below. First, the \texttt{System Stats Controller} initiates the I/O control by periodically collecting statistics from the \texttt{Lustre Job Stats Tracker} (\circled{1}) in the OST and pre-processing them before calling the \texttt{Token Allocation Algorithm} (\circled{2}). The token allocation algorithm queries the \texttt{Job Records}, which are maintained by AdapTBF framework to track lending and borrowing activities per OST (\circled{3}), and then determines the token allocation accordingly. 
After calculating the allocation, the algorithm will also update the \texttt{Job Records} (\circled{4}) to reflect any change to the job's tokens. 
The calculated token allocations are then passed to the \texttt{Rule Management Daemon} (\circled{5}), which checks for existing rules (\circled{6}) and creates or stops rules if necessary, before applying the token rates (\circled{7}). The token rates are applied via Lustre's Token Bucket Filter mechanism. 
Once the rules are applied, the daemon notifies the \texttt{System Stats Controller} (\circled{8}), and the controller clears the stats collected by the job tracker (\circled{9}). This process repeats at a user-defined frequency, allowing the AdapTBF framework to automatically adjust the token allocations for jobs at runtime. More details about each relevant component and the token allocation algorithm (with borrowing/lending mechanism) will be described in later subsections. Note that, all the notions used in this section are described in Table~\ref{tab:notion}.

\subsection{System Stats Controller}

In AdapTBF, the system stats controller gathers information from the job stats tracker located in the OST to identify active jobs during a specific observation period and determine their I/O demand. This information is crucial for the token allocation algorithm, which uses it to compute the appropriate token allocation for each active job. Once the token allocation is completed and the rule management daemon signals the controller, the controller clears the job stats. This process ensures that the controller only collects relevant data for each observation period, maintaining accurate and up-to-date statistics for future allocations.

\subsection{Token Allocation Algorithm}

The token allocation algorithm is the key component of AdapTBF framework. It runs the algorithm independently on each shared storage target (i.e., OST). The algorithm autonomously and adaptively allocates tokens for each active job on the specific storage target using only local information.

\begin{table}[h!]
\centering
\caption{Summary of main notations.}
\renewcommand{\arraystretch}{1.4}  
\begin{tabular}{p{0.06\textwidth} p{0.38\textwidth}}  
\toprule
\textbf{Notation} & \textbf{Description} \\
\midrule
$S_i$ & Object Storage Target, $i$ \\
$T_i$ & Maximum Token Rate(tokens/sec) of $S_i$ \\
$\Delta t$ & Observation period ($\Delta t = t_k - t_{k-1}$) \\
$J_i^{\Delta t}$ & Active jobs on $S_i$ during observation period $\Delta t$ \\
$J_+^{\Delta t}$ & Positive record jobs on $S_i$ during observation period $\Delta t$ \\
$J_-^{\Delta t}$ & Negative record jobs on $S_i$ during observation period $\Delta t$ \\
$n_x^t$ & Number of nodes allocated to job $x$ at time $t$ \\
$p_x^t$ & Priority of job $x$ at time $t$ \\
$r_x^t$ & Record of job $x$ at time $t$ \\
$r_{x,RD}^t$ & Record of job $x$ after redistribution at time $t$ \\
$r_{x,RC}^t$ & Record of job $x$ after re-compensation at time $t$ \\
$d_x^t$ & Observed I/O demand of job $x$ at time $t$ \\
$u_x^t$ & Utilization score of job $x$ at time $t$ \\
$\alpha_x^t$ & Allocated tokens of job $x$ at time $t$ \\
$\alpha_{x,RD}^t$ & Allocated tokens of job $x$ after redistribution at time $t$ \\
$\alpha_{x,RC}^t$ & Allocated tokens of job $x$ after re-compensation at time $t$ \\
$\rho_x^t$ & Remainder tokens of job $x$ at time $t$ \\
\bottomrule
\end{tabular}
\label{tab:notion}
\end{table} 

The algorithm operates in three sequential steps during each observation period: \texttt{Initial Allocation}, \texttt{Token Redistribution}, and \texttt{Token Re-compensation}. These steps enable the algorithm to account for the priority, I/O demand, efficiency, and fairness of each job. We describe these three steps individually below.

\subsubsection{Priority-Based Initial Allocation}

To ensure that jobs with more computational resources receive higher I/O bandwidth, we first determine the initial allocation solely based on each job’s priority. A job’s priority is defined by the number of compute nodes it requests relative to other active jobs on the storage target, such as 30\% or 0.3. The rationale behind granting higher allocations to larger jobs is to prevent them from being blocked or slowed down due to I/O bottlenecks.

The algorithm allocates tokens only to active jobs during the observation period $\Delta t$ on the shared storage target $S_i$. Active jobs are defined as those that have submitted RPC requests to the shared storage target during the observation period, allowing us to allocate tokens only to I/O-active jobs. Let $J_i^{\Delta t}$ denote the set of active jobs on $S_i$ during $\Delta t$, and let $n_x^t$ represent the number of computing nodes allocated to job $x$ at time $t$. The priority of job $x$ at time $t$ is then defined as:
\begin{equation} 
    p_x^t = \frac{n_x^t}{\sum_{x \in J_i^{\Delta t}} n_x^t} \quad \text{for} \quad \forall x \in J_i^{\Delta t}
\end{equation}

Defining $T_i$ as the maximum token rate (tokens/s) at $S_i$, the tokens allocated to job $x$ for the next period $\Delta t$, according to its priority $p_x^t$, is given by:
\begin{equation}
    \alpha_x^t = T_i \cdot p_x^t \cdot \Delta t \quad \text{for} \quad \forall x \in J_i^{\Delta t}
\end{equation}

\subsubsection{Redistribution of Surplus Tokens}
\label{subsec:redistribution}

The first step, initial allocation based on job priority, ensures that higher-priority jobs receive a larger proportion of the limited tokens. However, the job may not have that high I/O demand during that time interval. This discrepancy between allocated tokens and actual I/O demand can result in \textit{surplus} tokens that were allocated but remain unused, leading to system under-utilization as other jobs can not effectively utilize them. At the same time, jobs with high I/O demand but low priority may suffer from poor I/O performance due to their low token rate, even when unused tokens are available from other jobs. To mitigate this issue, we introduce the concept of \textit{surplus} tokens, and calculate it based on the observed I/O demand of each job. We propose to redistribute the total surplus tokens to all active jobs by considering their token utilization efficiency and priority as described in more details below.

First, we define the I/O demand of job $x$ at time $t$ as $d_x^t$, measured by the number of RPCs issued to the shared storage target $S_i$ during the observation period $\Delta t$ by job $x$. To assess the efficiency of token utilization for each job, we calculate the utilization score $u_x^t$ using the observed I/O demand and the allocated tokens from the previous time step (i.e., $\alpha_x^{t-1}$). This score provides insight into how effectively the job utilized its allocated tokens during the observation period:
\begin{equation}
    u_x^t = \frac{d_x^t}{\alpha_x^{t-1}} \quad \text{for} \quad \forall x \in J_i^{\Delta t}
\end{equation}

If more tokens are allocated to each job than its observed I/O demand in the previous time step, the difference is considered surplus tokens. The surplus token $T_s^x$ for job $x$ is defined as:
\begin{equation} 
    T_s^x = \begin{cases} 
    \alpha_x^t - d_x^t & \text{if } \alpha_x^t > d_x^t, \\
    0 & \text{otherwise} 
    \end{cases} \quad \text{for} \quad \forall x \in J_i^{\Delta t}
\end{equation}

Then, the total surplus tokens available for redistribution, $T_s$, can be calculated as:
\begin{equation}
    T_s = \sum_{x \in J_i^{\Delta t}} T_s^x
\end{equation}

By definition, surplus tokens will likely not be used by the application. To maximize the storage efficiency, we can redistribute it to other active jobs. However, different jobs certainly have different requirements on additional tokens. To redistribute the surplus tokens adaptively, we calculate a distribution factor for each job. Jobs with a token deficit (i.e., $u_x^t > 1$, where their observed I/O demand exceeds their allocated tokens) will be prioritized. Among these jobs, further ranking is based on their priority relative to other jobs. This prioritization ensures that jobs with higher I/O demand and greater need receive more surplus tokens, thereby improving I/O performance and system utilization. For jobs without a deficit, the distribution factor is calculated using both utilization and priority. Similarly, this ensures that jobs with better utilization and higher priority always receive a larger proportion of surplus tokens, so they are well prepared for future burst I/Os. The distribution factor, $DF_x^t$ for job $x$ at time $t$, is computed as:
\begin{equation}
    DF_x^t = \begin{cases}
    u_x^t + u_x^t \cdot p_x^t & \text{if } u_x^t > 1 \\
    u_x^t \cdot p_x^t & \text{if } u_x^t \leq 1
    \end{cases} \quad \text{for} \quad \forall x \in J_i^{\Delta t}
\end{equation}

Finally, the surplus tokens are redistributed among all jobs based on their share of the distribution factor. Both allocated tokens and record of all jobs are updated to reflect removed surplus tokens picked for redistribution and received additional tokens from redistribution:
\begin{equation}
    \alpha_{x,RD}^t = \alpha_x^t - T_s^x + \frac{DF_x^t}{\sum_{x \in J_i^{\Delta t}} DF_x^t} \cdot T_s \quad \text{for} \quad \forall x \in J_i^{\Delta t}
\end{equation}
\begin{equation}
    r_{x,RD}^t = r_x^t + T_s^x - \frac{DF_x^t}{\sum_{x \in J_i^{\Delta t}} DF_x^t} \cdot T_s \quad \text{for} \quad \forall x \in J_i^{\Delta t}
\end{equation}

AdapTBF maintains a record, $r_x$ for each job, which tracks tokens lent or borrowed. A positive record value indicates tokens lent by the job, while a negative value reflects tokens borrowed. Although tokens should ideally match job priority, the AdapTBF framework permits token lending and borrowing to enhance I/O efficiency and system utilization. To ensure fairness, borrowed tokens must be returned when needed.

\subsubsection{Re-compensation for Borrowed Tokens}
\label{subsec:recompensation}

The re-compensation step follows the token redistribution step to ensure that jobs that have lent tokens in previous step can be adaptively compensated when their I/O demands increase. This process reclaims tokens from jobs that have previously borrowed tokens.

In the AdapTBF framework, the reclaim process considers future estimated utilization, current utilization, and the priority of jobs involved in token lending to determine the amount of tokens to be reclaimed. Additionally, the recompensation is bounded by the borrowing record of the jobs from which tokens are reclaimed, ensuring that jobs are not overcompensated or throttled unnecessarily. This approach ensures fairness for lending jobs, helping them maintain higher efficiency and better resource utilization.

Jobs eligible for recompensation, denoted as $J_{+}^{\Delta t}$, are those with positive records. Conversely, jobs with negative records, $J_{-}^{\Delta t}$, serve as potential sources for token reclaiming. We only include jobs in these two sets if they maintain a positive or negative record both before and after redistribution, ensuring that the re-compensation process remains fair and accurate.
\begin{equation}
    J_{+}^{\Delta t} = \{x : x \in J_i^{\Delta t} \text{ and } r_x^t > 0 \text{ and } r_{x,RD}^t > 0\}
\end{equation}
\begin{equation}
    J_{-}^{\Delta t} = \{x : x \in J_i^{\Delta t} \text{ and } r_x^t < 0 \text{ and } r_{x,RD}^t < 0\}
\end{equation}

To prevent overcompensation, we must carefully determine what portions of allocated tokens from negative record jobs can be reclaimed by jobs with positive records. This involves considering several factors, including how much of a positive record job’s future token demand has already been fulfilled by its current allocation. We estimate the future utilization score $\bar{u}_x^{t+\Delta t}$ as:
\begin{equation}
    \bar{u}_{x}^{(t+\Delta t)} = \frac{\bar{d}_x^{(t+\Delta t)}}{\alpha_{x,RD}^t} \quad \text{for} \quad \forall x \in J_{+}^{\Delta t}
\end{equation}

where $\bar{d}_x^{(t+\Delta t)}$ is the estimated I/O demand for job $x$ at time $t+\Delta t$ and $\alpha_{x,RD}^t$ is the amount of tokens allocated to job $x$ after redistribution at time $t$. We assume $\bar{d}_x^{(t+\Delta t)} = d_x^t$, meaning the I/O demand will remain consistent in the next time step. Even if the assumptions miss for time $t$, AdapTBF will still catch up in the next time window, as shown in later evaluations. The future utilization score simplifies to:

\begin{equation}
    \bar{u}_{x}^{(t+\Delta t)} = \frac{d_x^t}{\alpha_{x,RD}^t} \quad \text{for} \quad \forall x \in J_{+}^{\Delta t}
\end{equation}

If the estimated future utilization score is high based on current token allocations, the positive record job should reclaim fewer tokens. On the other hand, if the current utilization score is high based on previous token allocations, the positive record job should reclaim more, similar to if its priority is also high. The reasoning is that jobs with higher current utilization and priority are likely to exhibit higher future I/O demand. The reclaim coefficient, $C_x^t$, for job $x$ at time $t$, determining the portion of tokens to be reclaimed, is computed as:
\begin{equation}
    C_{x}^t = \sum_{x \in J_{+}^{\Delta t}} p_x^t \cdot \frac{ \max(1, u_x^t) + \max(0, (1 - \bar{u}_{x}^{(t+\Delta t)}))}{2}
\end{equation}

The reclaimable amount from each negative record job is bounded by the magnitude of its negative record (i.e., the borrowed tokens). The final reclaim amount is computed as:
\begin{equation}
    T_R^x = \min(|r_x^t|, \left\lfloor C_{x}^t \cdot \alpha_{x,RD}^t \right\rfloor) \quad \text{for} \quad \forall x \in J_{-}^{\Delta t}
\end{equation}

Both the allocated tokens and the job’s record are updated to reflect tokens removed during re-compensation:
\begin{equation}
    \alpha_{x,RC}^t = \alpha_{x,RD}^t - T_R^x \quad \text{for} \quad \forall x \in J_{-}^{\Delta t}
\end{equation}
\begin{equation}
    r_{x,RC}^t = r_{x,RD}^t + T_R^x \quad \text{for} \quad \forall x \in J_{-}^{\Delta t}
\end{equation}

The total tokens available for recompensation, $T_R$, are calculated as:
\begin{equation}
    T_R = \sum_{x \in J_{-}^{\Delta t}} T_R^x
\end{equation}

For each job $x$, the recompensation factor $RF_x^t$ is the same as the redistribution factor $DF_x^t$. This equivalence ensures similar prioritization for jobs in $J_{+}^{\Delta t}$—those with a deficit and higher priority are compensated first, followed by jobs with no deficit, prioritized by utilization and priority:
\begin{equation}
    RF_x^t = DF_x^t \quad \text{for} \quad \forall x \in J_{+}^{\Delta t}
\end{equation}

Finally, the reclaimed tokens are redistributed among all jobs in $J_{+}^{\Delta t}$ based on their share of the distribution factor. The redistributed tokens update both the allocated tokens and the job’s record:

\begin{equation}
    \alpha_{x,RC}^t = \alpha_{x,RD}^t + \frac{DF_x^t}{\sum_{x \in J_{+}^{\Delta t}} DF_x^t} \cdot T_R \quad \text{for} \quad \forall x \in J_{+}^{\Delta t}
\end{equation}
\begin{equation}
    r_{x,RC}^t = r_{x,RD}^t - \frac{DF_x^t}{\sum_{x \in J_{+}^{\Delta t}} DF_x^t} \cdot T_R \quad \text{for} \quad \forall x \in J_{+}^{\Delta t}
\end{equation}

\subsubsection{Ensuring Fairness with Remainders}

To address the issue of integer division and ensure long-term fairness, we accumulate fractional remainders when distributing tokens at each step (e.g., priority allocation, surplus allocation, reclaim allocation). During each allocation step, any fractional tokens that cannot be allocated due to the integer constraint are stored as remainders for each job. In subsequent iterations, these remainders are added to the raw allocation before flooring, ensuring that jobs receive their fair share over time. If we represent the allocations and remainders for a job $x$ over multiple steps as a series $\alpha_{x,st}$ and $\rho_{x,st}$, where all jobs served by $S_i$ are denoted by $J$, we have:
\begin{equation}
    \alpha_{x,st} = \left[\alpha_x^0, \dots, \alpha_x^t, \alpha_{x,RD}^t, \alpha_{x,RC}^t, \alpha_x^{t+{\Delta t}}, \dots \right] \quad \forall x \in J
\end{equation}
\begin{equation}
    \rho_{x,st} = \left[\rho_x^0, \dots, \rho_x^t, \rho_{x,RD}^t, \rho_{x,RC}^t, \rho_x^{t+{\Delta t}}, \dots \right] \quad \forall x \in J
\end{equation}

The floored allocated tokens for job $x$ at step $j$, considering remainders, are given by:
\begin{equation}
    \alpha_{x,st}^f = \left\lfloor \alpha_{x,st}[j] + \rho_{x,st}[j-1] \right\rfloor \quad \text{for} \quad \forall x \in J_{i}^{\Delta t}
\end{equation}

where $\rho_{x,st}[j-1]$ represents the remainder from the previous allocation step. The new remainder and allocated tokens are updated as:
\begin{equation}
    \rho_{x,st}[j] = \alpha_{x,st}[j] - \alpha_{x,st}^f  \quad \text{for} \quad \forall x \in J_{i}^{\Delta t}
\end{equation}
\begin{equation}
    \alpha_{x,st}[j] = \alpha_{x,st}^f  \quad \text{for} \quad \forall x \in J_{i}^{\Delta t}
\end{equation}

However, adjusting the allocated tokens with the remainder can lead to scenarios where excess tokens are allocated or leftover tokens remain, both of which violate the total tokens to distribute constraint at the allocation step. To address this, we follow the largest remainder first fairness approach. In this method, we either reduce allocated tokens by one (in the case of excess) or increase allocated tokens by one (in the case of leftover) for the job with the largest remainder first. As we update the allocated tokens, we also adjust the remainder. This process is repeated until the total tokens meet the distribution constraint.

\subsection{Rule Management Daemon}

The TBF mechanism enforces bandwidth control by creating rules for specific classes, which can be defined based on \texttt{NID}, \texttt{JobID}, \texttt{Opcode}, and other parameters as required by Lustre. In the AdapTBF framework, we use \texttt{JobID} to classify incoming RPCs into these queues. The TBF rules ensure that jobs adhere to the set token rate when making RPC requests to the server, maintaining bandwidth control. The TBF mechanism supports dynamic rule creation and modification and allows the establishment of a hierarchical structure among rules.

The Rule Management Daemon interacts directly with the Lustre TBF mechanism to apply TBF rules for each active job, ensuring that the allocated tokens, as determined by the token allocation algorithm, are reflected in the active rules. Upon receiving a list of jobs with their allocated tokens, the daemon first checks the current TBF rules to determine if any rules for inactive jobs need to be stopped or if new rules need to be created for active jobs without existing rules. Importantly, jobs without rules do not suffer from starvation, as Lustre maintains a fallback queue for all incoming RPCs that do not match any rule. The fallback queue operates without a token rate limit, allowing RPCs to be served if any idle I/O threads are available.

The daemon then apply the token rate for each job based on the calculated allocation. During this process, the daemon establishes a hierarchy among the rules according to job priority, enabling idle threads to prioritize high-priority job queues if RPCs are present. Once the update is complete, the daemon notifies the System Stats Controller that the rule management has completed. In this way, the Rule Management Daemon dynamically manages TBF rules, ensuring efficient enforcement of token rate updates.

\subsection{Generalization of AdapTBF}
{AdapTBF framework provides decentralized automated bandwidth control of individual storage targets (i.e. OSTs). Although it is tied to Lustre due to its reliance on the mature TBF mechanism implemented in Lustre, our key contribution, the adaptive token-borrowing mechanism, can be general to other distributed storage systems or even other scenarios. Specifically, it can be applied to situations involving the adaptive allocation of shared, finite resources among competing entities in a decentralized manner.
}
\section{Evaluation}
\label{sec:evaluation}

\subsection{Evaluation Platform}

We utilized the CloudLab~\cite{duplyakin2019design} platform to conduct our evaluations. We chose CloudLab because it is an open platform, making all reported results reproducible for other researchers. We deployed Lustre~\cite{braam2019lustre} version 2.15.5 as the parallel file system, using six CloudLab c6525-25g machines, as detailed in Table~\ref{tab:hardware_spec}, to set up a prototype cluster. The cluster configuration included one machine serving as both the Management Server and Metadata Server, one machine as the Object Storage Server (OSS) with two Object Storage Targets (OSTs), and four machines designated as Lustre clients.

\begin{table}[htpb]
\footnotesize
\centering
\caption{Hardware Specification for c6525-25g Node}
\begin{tabular}{ll}
\toprule
\textbf{Component} & \textbf{Specification} \\ 
\midrule
CPU   & 16-core AMD 7302P at 3.00GHz \\ 
\midrule
RAM   & 128GB ECC Memory (8x 16 GB 3200MT/s RDIMMs) \\ 
\midrule
Disk  & Two 480 GB 6G SATA SSD \\ 
\midrule
NIC   & Two dual-port 25Gb GB NIC (PCIe v4.0) \\ 
\bottomrule
\end{tabular}
\label{tab:hardware_spec}
\end{table}

\subsection{AdapTBF Implementation}
The proposed AdapTBF framework is currently implemented based on Lustre TBF mechanism. 
For collecting job information, AdapTBF queries the \texttt{job\_stats} of the respective Lustre OST. The framework manages TBF rules, such as creating, modifying, or stopping them, by leveraging \texttt{nrs\_tbf\_rule}, which is available in the Lustre OSS module. In our implementation, we set \texttt{jobid\_var} to $nodelocal$ and \texttt{jobid\_name} to $\%e.\%H$ to generate unique job IDs for the running applications.

\subsection{Evaluation Baselines}
\label{subsec:baseline}

We evaluated the AdapTBF framework to demonstrate its ability to improve throughput, adjust to varying I/O loads, enhance utilization, and ensure fairness. We compared AdapTBF against two bandwidth (BW) control scenarios: \texttt{No BW} and \texttt{Static BW}. In the No BW scenario, there is no bandwidth control applied to the OST, meaning no TBF rules are used to regulate the RPC stream in the OSS, which represents the default environment in the Lustre file system. The RPCs get served by the available I/O thread in First-Come First-Serve (FCFS) manner. In the Static BW scenario, static TBF rules are applied to the OSS for the OST based on the calculated priority, determined by the proportion of allocated resources relative to the total resources available in the system.

Although many bandwidth-sharing strategies exist (summarized in this survey~\cite{benoit2024revisiting}), most are not directly comparable to our framework due to their lack of priority awareness, dependence on I/O patterns, or inability to adapt bandwidth allocation fairly. So, we did not include them in our comparisons as it would be unfair.
The most comparable is GIFT~\cite{patel2020gift}, which centrally allocates bandwidth and throttles applications to maintain target bandwidth, rewarding throttled ones with coupons. In contrast, AdapTBF avoids throttling applications that use their allocated bandwidth within priority limits, ensuring adaptive and fair resource allocation without hindering progress.
Additionally, GIFT’s throttling strategy overlooks application priorities, potentially hindering high-priority jobs, and its centralized control adds significant overhead, making it unsuitable for large-scale HPC systems. For these reasons, we did not include evaluations of GIFT. Since AdapTBF is essentially an improvement over basic TBF mechanism, we focus on conducting extensive evaluations to show its advantages in this study.

\begin{figure}[htpb]
\begin{center}
\includegraphics[width=0.48\textwidth]{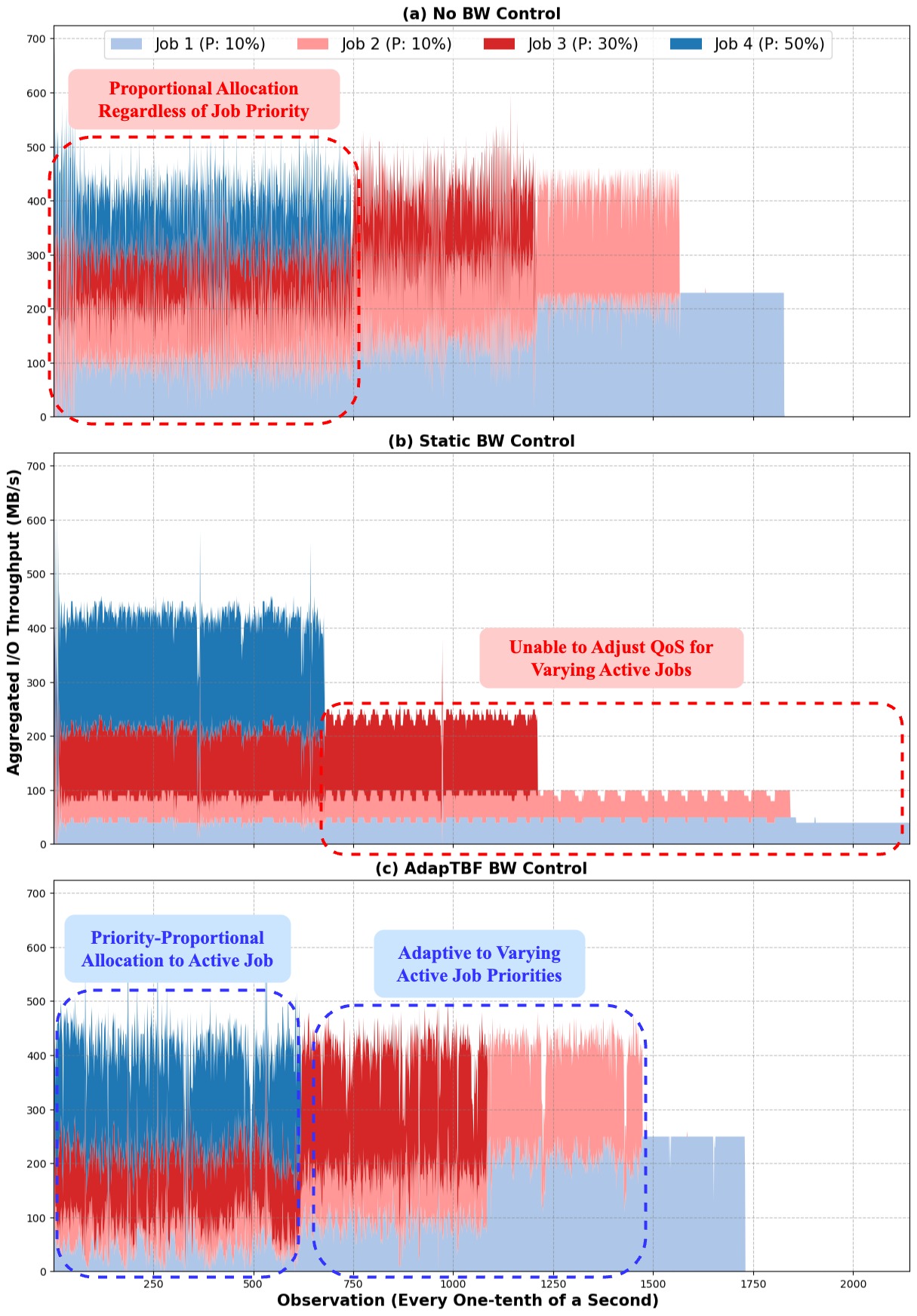}
\caption{Analysis of Jobs Execution in Section~\ref{subsec:prio}}
\label{fig:sc1_analysis}
\end{center}
\end{figure}

\subsection{Evaluation on Token Allocation}
\label{subsec:prio}

\subsubsection{Evaluation Goal}
The goal of this evaluation is to demonstrate the \textit{AdapTBF framework's ability to adjust token allocation proportional to job priority and to adapt to the changing set of active jobs}. 
When all jobs are I/O-intensive, it is crucial to prioritize high-priority jobs. When jobs complete execution, we must adjust token allocation to the remaining active jobs, ensuring maximum system resource utilization.

\subsubsection{Jobs I/O Pattern}
To simulate this scenario, we manually created four jobs ($Job_1-Job_4$), with the same I/O pattern and client configuration but different job priorities ($Job_1=10\%, Job_2=10\%, Job_3=30\%, Job_4=50\%$). All job patterns in the evaluations were generated using the Filebench benchmark~\cite{tarasov2016filebench}. Each job runs 16 processes, performing sequential I/O to its own unique file in a file-per-process manner, with each file sized at 1 GB. Note that, we manually set the priority of each job to evaluate AdapTBF. In addition, the job execution time also varies with higher-priority jobs finishing earlier, simulating a dynamic set of active jobs.


\subsubsection{Evaluation Results Analysis}
Figure~\ref{fig:sc1_analysis} reports the I/O bandwidth timeline of all four jobs under different I/O bandwidth control mechanisms. The Y-axis represents the aggregated I/O throughput achieved by four jobs on the OST and the X-axis represents observation collected at every $100 ms$. The axis representation remains consistent for rest plots.

\begin{figure}[htpb]
\begin{center}
\includegraphics[width=0.48\textwidth]{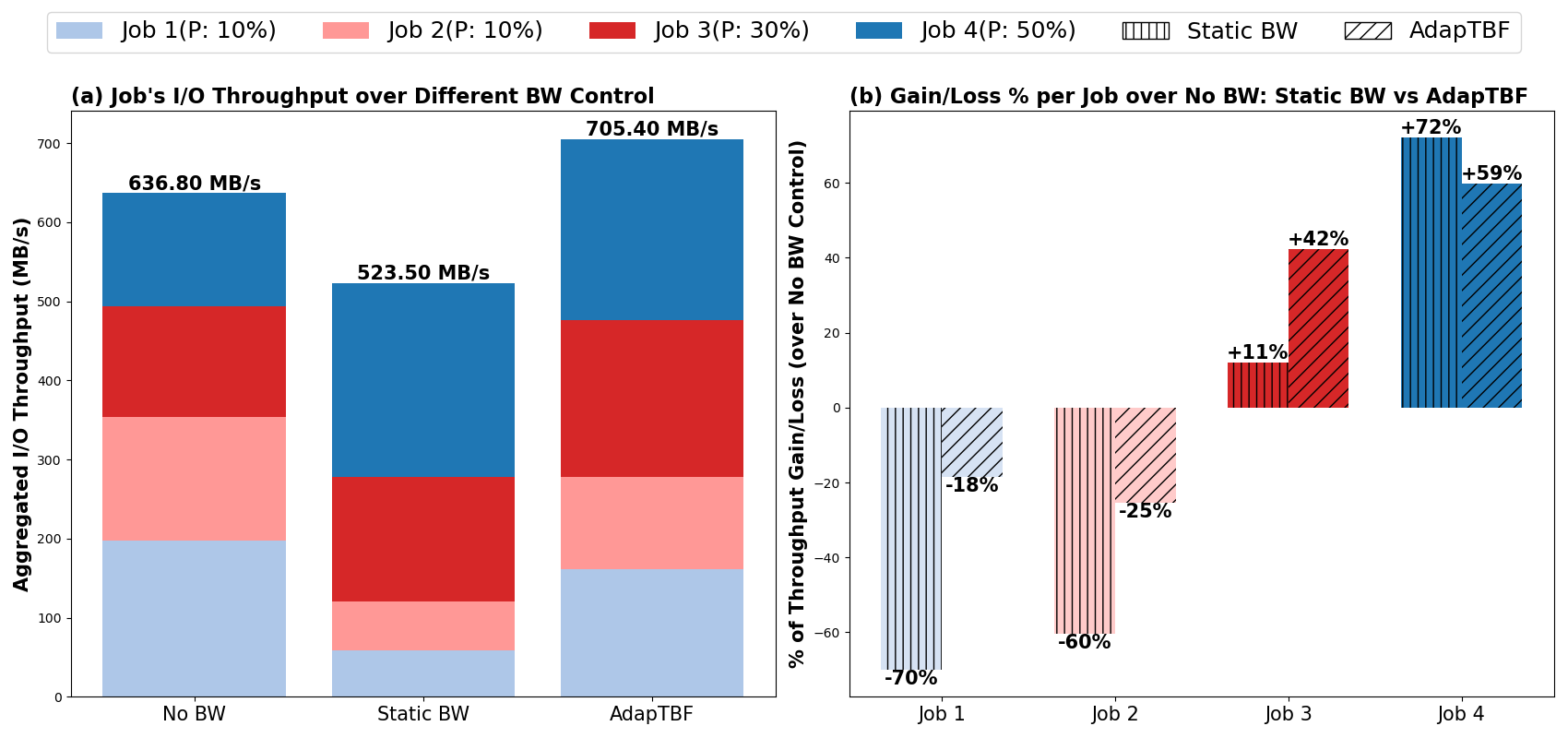}
\caption{Results of Jobs Execution in Section~\ref{subsec:prio}}
\label{fig:sc1_result}
\end{center}
\end{figure}

As shown in Figure~\ref{fig:sc1_analysis}(c), the AdapTBF framework allocates tokens proportional to job priority, in contrast to No BW control, which does not follow priority-proportional allocation (Figure~\ref{fig:sc1_analysis}(a)). Additionally, AdapTBF adjusts token allocation at each observation step based on the active jobs, unlike Static BW control, which does not adapt to changing job sets (Figure~\ref{fig:sc1_analysis}(b)). 
In Figure~\ref{fig:sc1_result}(a), we further show the achieved I/O bandwidth per job as well as overall under different mechanisms. AdapTBF is able to gain the highest overall I/O throughput as well as distributing more bandwidth to higher priority jobs ($Job_3$ and $Job_4$). Figure~\ref{fig:sc1_result}(b) details the I/O bandwidth gains/losses due to AdapTBF comparing with no bandwidth control case. We can observe that AdapTBF achieves significant throughput gains for $Job_3$ and $Job_4$, which have higher priority. Meanwhile, $Job_1$ and $Job_2$, which receive fewer tokens due to their lower priority, experience minimal throughput loss due to the framework's ability to adapt to the active set of jobs and their utilization.

\subsection{Evaluation on Token Redistribution}
\label{subsec:adap1}

\subsubsection{Evaluation Goal}
The motivation for this evaluation is to assess the \textit{AdapTBF framework’s ability to redistribute tokens} in an extreme scenario, where high-priority jobs generate short bursts of I/O, while low-priority jobs are I/O-intensive with continuous I/O. 
Ideally, an adaptive I/O control framework should allow low-priority jobs to utilize I/O resources while high-priority jobs are not, creating periods of I/O under-utilization. When a high-priority job experiences an I/O burst, the system should quickly process this burst to meet the job's bandwidth requirements.

\subsubsection{Jobs I/O Pattern}
To simulate this scenario, we run four jobs ($Job_1$ - $Job_4$). The first three jobs have high priority ($30\%$ each) and generate periodic short I/O bursts with varying burst volumes and intervals. The fourth job ($Job_4$) has a low priority ($10\%$) but exhibits a high, continuous I/O demand. The first three jobs each run 2 processes, performing sequential I/O in a file-per-process manner, with each file sized at 1 GB. These jobs issue I/O requests in a periodic manner, but the intervals and burst magnitudes are varied among jobs to create an interleaved pattern of I/O bursts on the server. $Job_4$ runs 16 processes, also performing sequential I/O in a file-per-process manner with 1 GB files, but its I/O demand is continuous.

\begin{figure}[htpb]
\begin{center}
\includegraphics[width=0.48\textwidth]{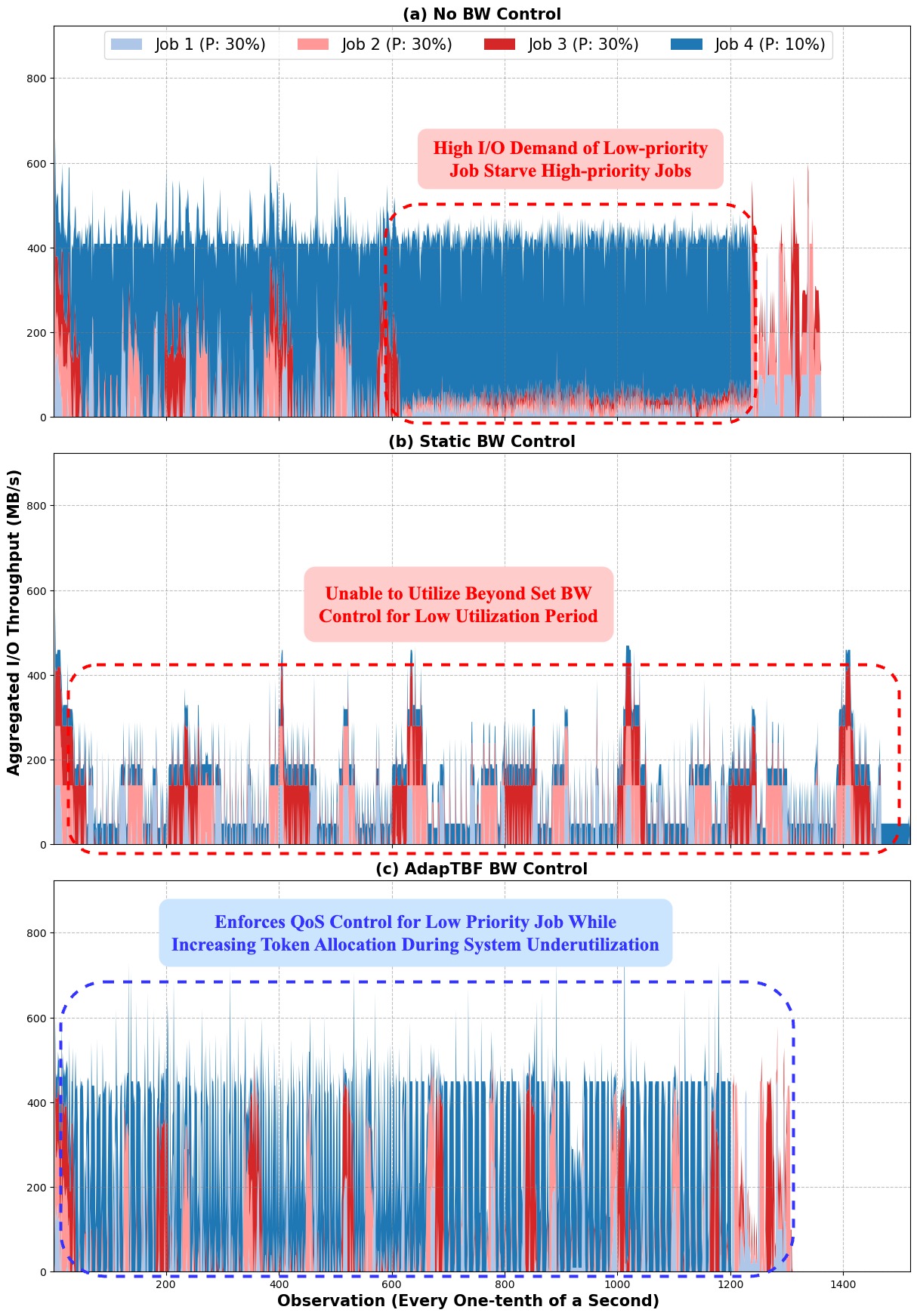}
\caption{Analysis of Jobs Execution in Section~\ref{subsec:adap1}}
\label{fig:sc2_analysis}
\end{center}
\end{figure}


\subsubsection{Evaluation Results Analysis}
\begin{figure}[htpb]
\begin{center}
\includegraphics[width=0.48\textwidth]{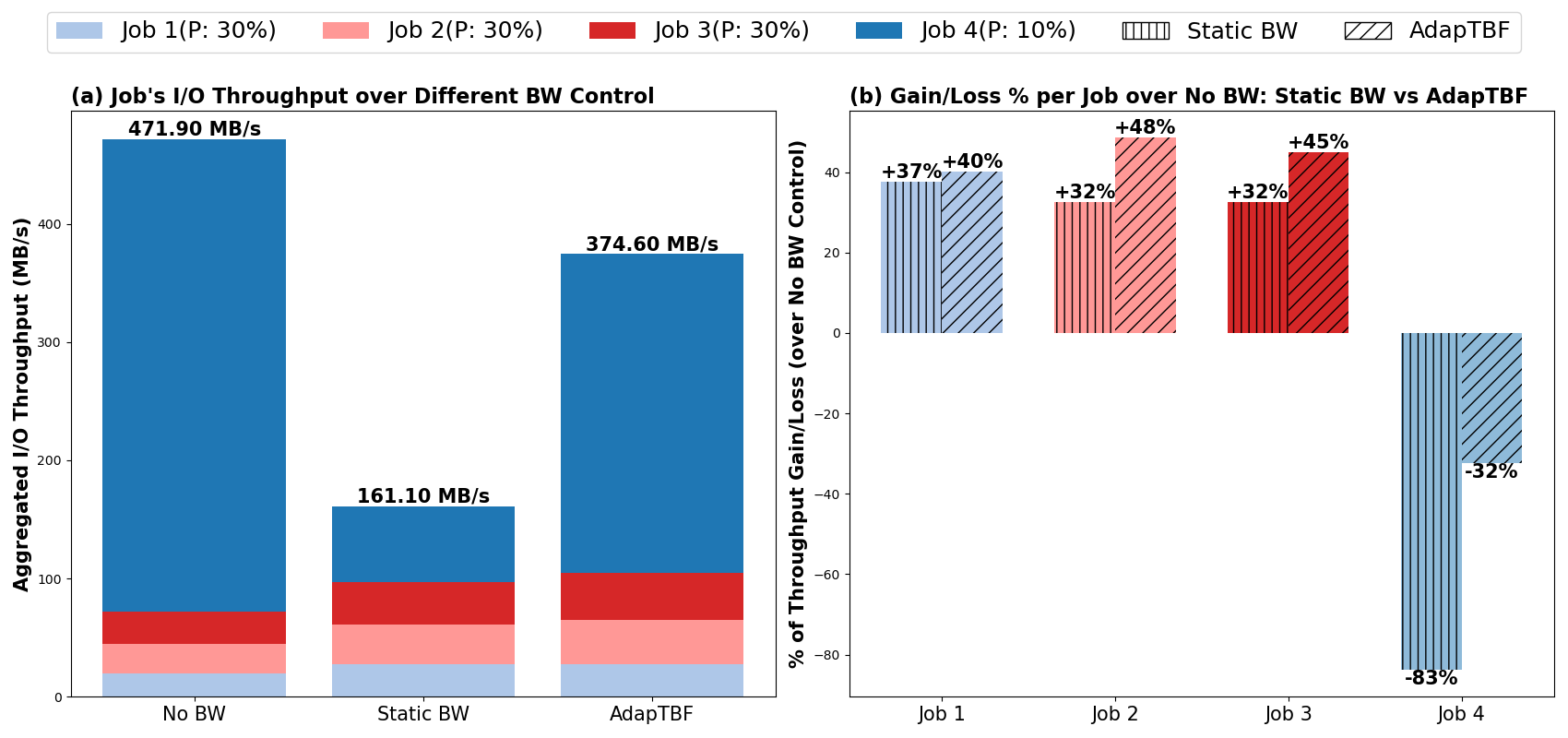}
\caption{Results of Jobs Execution in Section~\ref{subsec:adap1}}
\label{fig:sc2_result}
\end{center}
\end{figure}

Figure~\ref{fig:sc2_analysis} shows the results of different bandwidth control approaches.
As shown in Figure~\ref{fig:sc2_analysis}(c), the AdapTBF framework allocates tokens proportionally to both job priority and the varying I/O demand of the jobs. In contrast to No BW control (Figure~\ref{fig:sc2_analysis}(a)), which severely starves the high-priority jobs (both before and after $Job_4$'s I/O burst) due to the high I/O demand from $Job_4$, the AdapTBF framework prevents such I/O interference from occurring. Additionally, AdapTBF adjusts token allocation at each observation step based on the utilization of tokens by jobs, unlike Static BW control, which does not adapt to changes in token utilization, leaving system resources underutilized (Figure~\ref{fig:sc2_analysis}(b)). In Figure~\ref{fig:sc2_result}(b), we observe that AdapTBF achieves significant throughput gains for Jobs 1-3 (high-priority jobs) compared to both No BW and Static BW. However, for $Job_4$ (and the aggregate throughput), AdapTBF limits its throughput due to its low priority.

The primary reason for this is that in the No BW scenario, $Job_4$ drives continuous RPC stream, leading to maximized system utilization but at the price of higher-priority jobs being starved. In contrast, the AdapTBF framework is able to assign a larger share of tokens (via lending/borrowing) to high-priority jobs to handle their short I/O bursts (when happens), and compensate these tokens when when they do not fully utilize the allocated tokens. Nevertheless, AdapTBF quickly adjusts its allocation based on token utilization, minimizing the throughput loss for $Job_4$ while maintaining higher throughput for the high-priority jobs, even when compared to a static high-proportional allocation as seen in Static BW scenario.

In Figure~\ref{fig:sc2_analysis}(c), where AdapTBF does not fully utilize the available tokens, we cannot simply allocate all unused tokens to $Job_4$ as we assume no knowledge of the job’s I/O pattern. This design choice keeps the AdapTBF framework agnostic to job-specific patterns. The file system could incorporate hints from the job tracker to identify I/O patterns, enabling AdapTBF to make more informed allocations. However, this enhancement is beyond the scope of the current study.

\begin{figure*}[htpb]
\begin{center}
\includegraphics[width=0.85\linewidth,height=8cm]{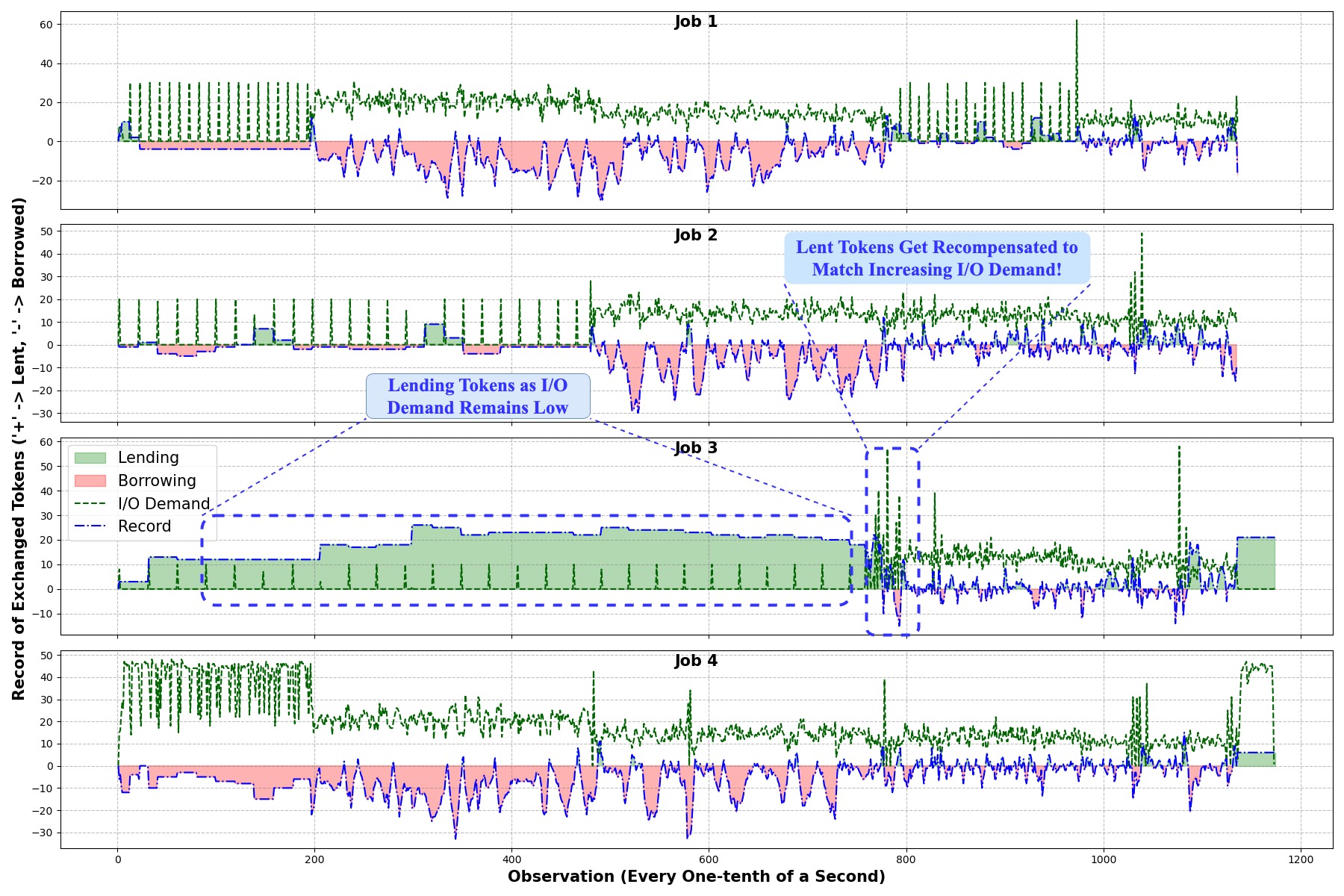}
\caption{Analysis of Jobs Execution in Section~\ref{subsec:adap2}}
\label{fig:sc3_analysis}
\end{center}
\end{figure*}

\subsection{Evaluation on Token Re-compensation}
\label{subsec:adap2}

\subsubsection{Evaluation Goal}
The goal of this evaluation is to assess the \textit{AdapTBF framework’s ability to re-compensate borrowed tokens to jobs that have lent them}. 
When a job has a low I/O load, the framework allows other jobs with high I/O demand to borrow tokens from that job, ensuring maximum utilization of limited system resources. However, if the I/O load of the lending job increases, the framework should re-compensate the job for the tokens it lent, in addition to its regular token allocation based on priority. This mechanism ensures the framework meets both the system’s utilization goals and the fairness objectives for all jobs.

\subsubsection{Jobs I/O Pattern}

To simulate this scenario, we run four jobs ($Job_1$ - $Job_4$) again. But this time, all of them have equal priority ($25\%$ each). The first three jobs each run 2 processes, performing sequential I/O in a file-per-process manner, with each file sized at 1 GB. One process from each of these three jobs issues I/O requests in small bursts at constant intervals, but the intervals and burst magnitudes vary among the jobs to create an interleaved pattern of I/O bursts on the server. The second process from each of these jobs starts issuing continuous I/O requests after a delay from the start of execution, with delays of 20s, 50s, and 80s for $Job_1$, $Job_2$, and $Job_3$, respectively. $Job_4$ runs 16 processes, also performing sequential I/O in a file-per-process manner with 1 GB files, but its I/O demand starts from the beginning and continous.


\subsubsection{Evaluation Results Analysis}
Figure~\ref{fig:sc3_analysis} shows trend of both record (i.e. exchanged count of tokens up to that observation) and I/O demand (i.e. number of RPCs, $1 RPC=1 Token$) per job in each subplot. As seen in Figure~\ref{fig:sc3_analysis}, $Job_3$, which has the largest delay and the smallest I/O burst among the first three jobs, lends its tokens during the first 80 seconds of execution. However, once the process of $Job_3$ that performs continuous I/O starts, the AdapTBF framework quickly begins re-compensating tokens from the jobs that previously borrowed tokens, as $Job_3$ exhibits a higher I/O load on the server. This demonstrates how efficiently the framework handles the re-compensation process in response to increased I/O demand, ensuring fairness for the jobs participating in the lending-borrowing cycle over the long term.

\begin{figure}[htpb]
\begin{center}
\includegraphics[width=0.48\textwidth]{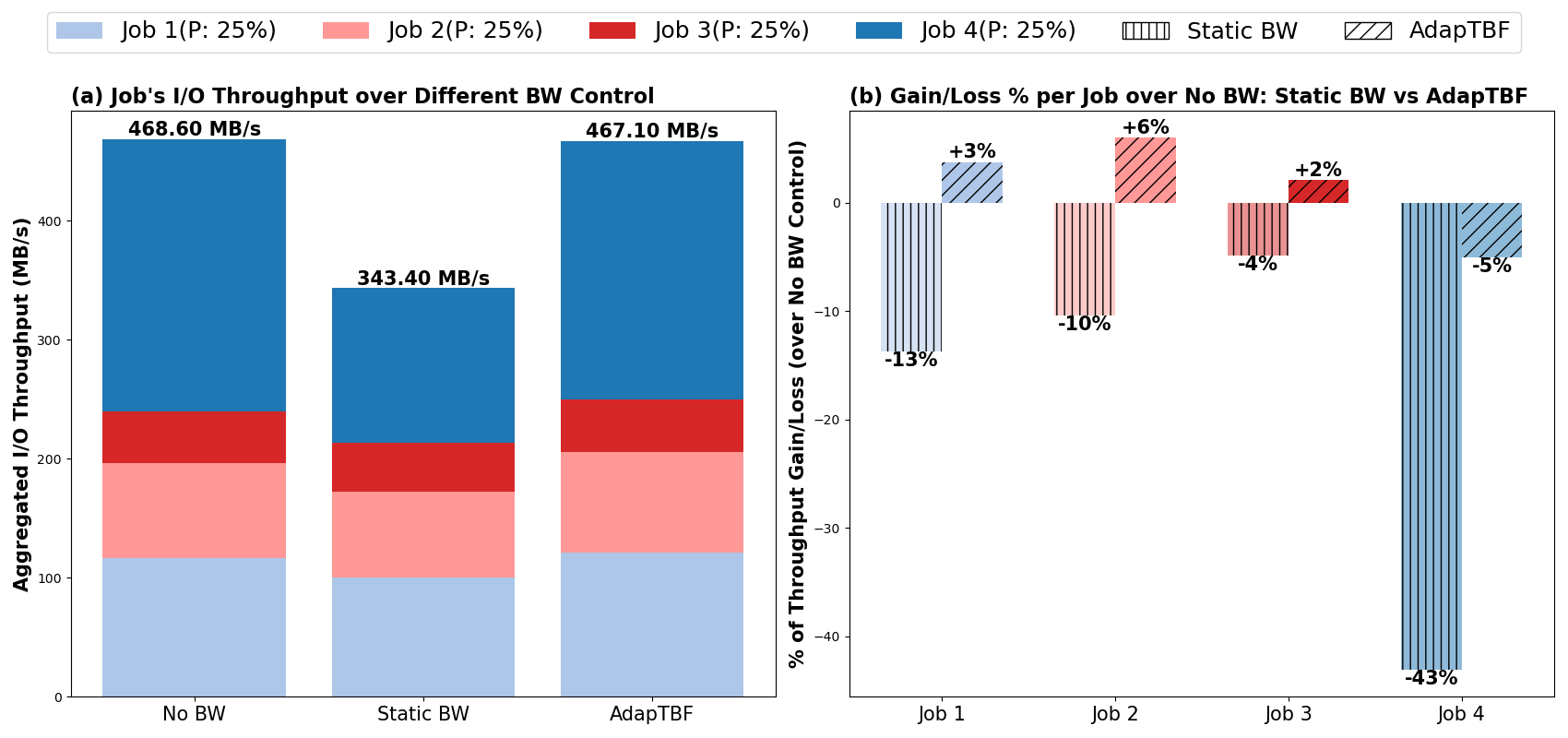}
\caption{Results of Jobs Execution in Section~\ref{subsec:adap2}}
\label{fig:sc3_result}
\end{center}
\end{figure}

In Figure~\ref{fig:sc3_result}(a), we observe that the AdapTBF framework performs on par with the No BW scenario, while the Static BW scenario suffers significant degradation. Furthermore, in Figure~\ref{fig:sc3_result}(b), AdapTBF achieves gains over both No BW and Static BW for the first three jobs, while incurring only minimal loss for $Job_4$ compared to the No BW scenario. The ability to achieve gains for bursty I/O while minimizing loss for continuous I/O demonstrates how quickly AdapTBF adapts to varying I/O loads by redistributing and recompensating tokens, as highlighted in Figure~\ref{fig:sc3_analysis}.

In contrast, Static BW maintains the same token allocation regardless of I/O load, which results in neither being able to adapt to I/O burst nor being able to fully utilize resources. Although the AdapTBF framework performed similarly to the No BW scenario in terms of aggregate I/O throughput, its emphasis on fairness is more evident. AdapTBF ensures that I/O bursts are addressed proportionally to job priority rather than purely based on I/O load observed by the server, highlighting its ability to balance fairness and performance.

\subsection{Analysis of AdapTBF Framework Overhead}

The AdapTBF framework operates independently across each OST, collecting information that is local to the specific OST. As the framework does not communicate or gather data outside of the designated OST, it incurs no external I/O and communication overhead, making it highly scalable to a large number of OSSes. This design ensures that the AdapTBF framework is well-suited for large-scale HPC environments.

The token allocation algorithm used by AdapTBF has a time complexity of $O(n)$, where $n$ represents the number of active jobs, resulting in linear scaling of allocation time with the number of active jobs. In our evaluation, the average time for token allocation was less than $30 \mu s$ per job, meaning that even with $1000$ active jobs, the calculation overhead would be less than $30ms$. The overall framework overhead, including the processes of collecting and clearing stats, as well as creating, modifying, and stopping rules, will not increase with the number of jobs. It was about $25ms$ in all our evaluation cases.

Integrating AdapTBF directly into Lustre would further reduce this overhead, as it would enable the framework to read, modify, and write information directly from Lustre's runtime memory, avoiding the additional overhead of external interactions. The memory footprint of AdapTBF is also minimal, as it only stores two pieces of information in runtime memory: the job ID and the value of the record.

\subsection{Evaluation on Token Allocation Frequency}

\begin{figure}[htpb]
\begin{center}
\includegraphics[width=0.48\textwidth]{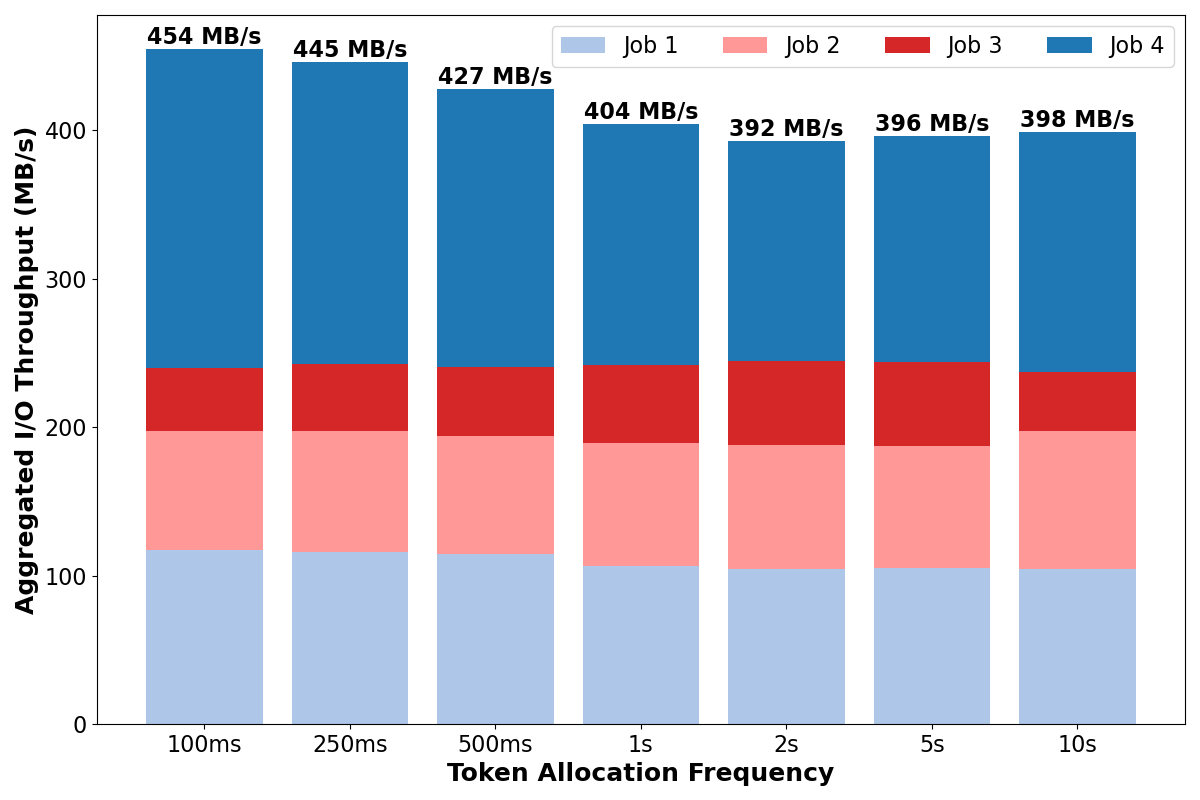}
\caption{I/O Throughput Comparison for Varying Token Allocation Frequency}
\label{fig:sc4_result}
\end{center}
\end{figure}

We performed an evaluation to understand the impact of token allocation frequency on the aggregate I/O throughput of jobs. The evaluation used the same job patterns and client configurations as described in Section~\ref{subsec:adap2}, chosen for their mix of small I/O bursts and continuous I/O streams. As shown in Figure~\ref{fig:sc4_result}, a smaller allocation frequency results in better I/O throughput, which aligns with the intuition that shorter intervals between consecutive allocation steps allow the framework to adapt faster and address sudden bursts more effectively. However, the allocation frequency cannot be set lower than the framework overhead. Given the low overhead of the AdapTBF framework, we selected 100 ms as the frequency for performing token allocation in all our experiments.

\section{Related Work}
\label{sec:related}

Enforcing bandwidth in shared storage has been intensively studied~\cite{huang2003stonehenge,wang2012cake,thereska2011sierra,gulati2009parda,wu2007providing,li2016pslo,qian2017configurable, van1998sleds,wong2006zygaria,lumb2003faccade,gildfind2011method,gulati2007pclock,gulati2010mclock,wachs2007argon,zhang2011qos,uttamchandani2004polus,sundaram2003practical,qian2013dynamic,xu2012vpfs,wang2007proportional,jin2004interposed}. The essential approach is common: first identify the I/O requests from different applications and utilize certain I/O scheduling algorithms to control which one should be served. The choices of I/O scheduling algorithms are huge. They can be based on fair queuing method such as Start-Time Fair Sharing (SFQ) and its many variations~\cite{goyal1996start}, or based on token method such as leaky token bucket and its variations~\cite{tang1999network}. AdapTBF follows the similar overall steps, but differs from existing studies for being a decentralized token bucket algorithm with token borrowing mechanism for work-conserving I/O control.

The most comparable strategy to AdapTBF is GIFT~\cite{patel2020gift}. We have discussed detailed comparisons to this work in Section~\ref{subsec:baseline}. Among other existing studies, there are three of them that are closely related to AdapTBF and worth discussing. 
Qian et al.~\cite{qian2017configurable} introduced QoS control into Lustre. It utilized token bucket filter (TBF) algorithm to support local I/O control and introduced a controller to globally monitor I/O activities on all storage servers and control token assignment accordingly. Wu et al.~\cite{wu2007providing} proposed an QoS framework for Ceph file system, which utilized weighted round robin (WRR) to schedule local I/Os. It assumes the workloads are evenly distributed to all OSDs to simplify the global I/O control. vPFS~\cite{xu2012vpfs} utilized Start-Time Fair Sharing with Depth (SFQ(D)) to share local I/O bandwidth among applications, and used the batch-based monitor-and-control method to control global I/O bandwidth.
AdapTBF shares some design insights with these studies, but significantly differs regarding how bandwidth control is implemented. Specifically, AdapTBF uses a decentralized token borrowing/lending mechanism to ensure tokens will be automatically adapted based on actual I/O demand. This leads to better adaptation to the workload changes of applications as our results show. 


The software-defined storage (SDS) is also relevant with AdapTBF as it also aims to control the I/O activities in large-scale storage systems. However, most of the SDS systems are built for end-to-end (e2e) policy enforcement and touches many layers of the storage stack in data centers. For example, IOFlow~\cite{thereska2013ioflow}, later extended as sRoute~\cite{stefanovici2016sroute}, provides the ability to track, schedule, and control the I/O path from VMs to shared storage. The goal of AdapTBF is not to enforce the SLA agreement to users, instead, it is for guarding the abnormal I/O behaviors of applications as well as maximizing the whole system utilization. In addition, the decentralized global I/O control of AdapTBF is also different from the centralized controller used in IOFlow and sRoute. Crystal~\cite{gracia2017crystal} goes beyond IOFlow and sRoute by providing richer data planes and different suites of management abstractions. However, it is not designed for HPC platforms where data and compute resources are separated.

{
HPC I/O tuning mechanisms, such as async I/O~\cite{tang2021transparent,tarraf2024behind} for overlapping computation with I/O, pattern-driven data placement~\cite{neuwirth2016using, paul2024tarazu,zhou2021characteristic} to reduce intra-application interference, and network-level mitigation strategies~\cite{smith2018mitigating, kang2023workload,zhou2019prs} for managing bursts, operate at different system components. These methods complement AdapTBF, which leverages native file system support to extend TBF at the OST level, efficiently managing RPCs and enforcing scalable I/O control.
}

\section{Conclusion and Future Work}
\label{sec:conclusion}

Highly concurrent data accesses on the shared HPC storage system can easily interfere with each other, leading to significant performance issues. It is critical for the storage systems to monitor and control applications' I/O activities and to guard necessary bandwidth for applications. In this research, we have introduced AdapTBF, a framework to adaptively control I/O bandwidth on the storage storage servers utilizing the well-established TBF mechanism. Through the decentralized token borrowing/lending mechanism, AdapTBF addresses the efficiency issue of naive token bucket mechanism. We believe that AdapTBF and its associated design could become a necessary component of future parallel file systems. Our future work will focus on integrating the AdapTBF into representative parallel file systems like Lustre. 


\section*{Acknowledgments}
We sincerely thank the anonymous reviewers for their valuable feedback. This work was supported in part by National Science Foundation (NSF) under grants CNS-2008265 and CCF-2412345,

\bibliographystyle{plain}
{
\bibliography{bib}
}
\end{document}